\documentclass{article}

\usepackage{amsmath}
\usepackage{amsfonts}
\usepackage{amssymb}
\usepackage{graphicx}
\usepackage{graphics}

\begin{document}

\title{THE STRUCTURE OF ULTRATHIN Ag FILMS ON Pd(111)}

\author{PREDRAG LAZI\' C, DAMIR \v SOK\v CEVI\' C and RADOVAN BRAKO\\
\textit{Rudjer Bo\v skovi\' c Institute, P.O. Box 180, HR-10002 Zagreb, Croatia}}

\maketitle

\begin{abstract}
\noindent{We have performed \textit{ab initio} density functional calculations of 
thin Ag films on the Pd(111) surface. We have calculated the
structural properties and the electronic bands of the Ag/Pd systems.
There is a band gap in the electronic density of states around the centre of the 
two-dimensional Brillouin zone of the Pd(111) surface, which makes possible
the formation of localised states in the adsorbed silver films. 
We find that quantum well states may form at binding energies around
4~eV.}\\

\noindent{PACS numbers: 71.15.Mb, 73.20.At, 73.21.Fg, 73.61.At} \\

\noindent{Keywords: density functional calculations, thin films, silver, palladium, 
quantum well states}
\end{abstract}

\section{Introduction}

Ultrathin metallic films with thickness ranging from a monolayer to a few tens of
layers deposited on low index surfaces of a different metal show a wide variety
of structural and electronic properties. The scanning tunnelling 
microscopy (STM) and the angle resolved photoemission spectroscopy (ARPES)
make it possible to closely monitor the growth of the films and the evolution of 
the electronic states~\cite{Milun02,Chiang00}. 
Depending on the difference of the 
bulk lattice constants of the two metals, the films grow either
in registry, where the atoms of the second species take the same positions
which would be occupied in further substrate growth, 
or form layers denser or more open
than the substrate often producing Moire
patterns in the STM images, or some form of alloying of the 
substrate and the deposited metal can occur. The electronic structure
can also show interesting properties, such as the formation of quantum
well (QW) states localised on the adlayers in cases
when there is an energy gap of the substrate states,
or a symmetry incompatibility between the substrate and 
adlayer states leads to an effective ``symmetry gap'' for the adlayer 
electrons.

Silver and palladium are
in many respects quite similar metals. Both have fcc structure,
with bulk lattice constants of 4.09 {\AA} and 3.89 \AA,
respectively. Experiments show that the silver films on Pd(111) surface
start to grow pseudomorphically~\cite{Eisenhut93}.

In this paper we report \textit{ab initio} density functional calculations of
one to three Ag layers on a Pd(111) surface. We calculate the structural
properties, such as the energies of formation of various structures, the 
equilibrium distances between layers, etc. We also use the Kohn-Sham 
wavefunctions obtained in the calculation to discuss the electronic
structure of the Ag/Pd(111) system, looking in particular into the 
formation of quantum well states localised around the centre of the 
surface Brillouin zone, which can be observed in normal ARPES.

\section{Calculation}

We have made the \textit{ab initio} density functional calculations using the
DACAPO program package~\cite{dacapo}. We used periodic boundary conditions in 
all three spatial
directions, modelling the Pd substrate by six hexagonal layers
and a vacuum region of sufficient thickness which separates the
periodically repeating Pd slabs.
The details of the calculations have been described in
papers dealing with similar systems~\cite{Kralj03}. The lattice constant of Pd was 
taken to be 3.99 \AA, which is the equilibrium value for our calculations
of bulk Pd, and differs slightly from the experimental value of 3.89 \AA.
This choice is necessary in order to avoid the appearance of spurious 
stress in the substrate. The bottom three Pd layers were kept fixed at bulk
separation, and the top three layers of the substrate
and the silver adlayers were allowed to relax freely in the direction
perpendicular to the surface ($z$ coordinate). 

\section{The structure and growth of Ag films}

It has been found experimentally that silver films grow in registry on the
Pd(111) substrate, but with a stacking fault between the first and the
second layer, resulting in a twinned crystalline structure~\cite{Eisenhut93}. 
We have therefore investigated several configurations of Ag layers on a Pd substrate.
First, we have taken the fcc palladium (in our calculation this is 
six hexagonal layers of atoms stacked in the fcc
ABCABC order) and added one to three layers of Ag continuing the
regular fcc ordering (e.g. ABCABC|AB) and allowing the relaxation.
Next, we have calculated
the structures in which a silver layer (in particular the second one)
goes into the hcp position, resulting in structures like ABCABC|AC, etc.
The calculated interlayer distances and adsorption energies are shown in 
Table~\ref{properties}. 

\renewcommand{\baselinestretch}{1.0}
\begin{table}[htb]
\caption{Step heights, interlayer distances
(both in \AA) and adsorption energies for Ag layers on Pd(111). 
\hfill\newline}
\label{properties}
\begin{tabular*}{\textwidth}{|c@{\extracolsep{\fill}}|cccc|} \hline
           &    Clean Pd  &  1 ML Ag    & 2 ML Ag    &   2 ML Ag w.  \\
           &              &             &            &  stacking fault \\ \hline
Step height&              &   2.37      & 2.34       &    2.44 \\ \hline
Ag$_1$--Ag$_2$&           &             & 2.42       &    2.45  \\
Pd$_1$--Ag$_1$&           &   2.36      & 2.34       &    2.36  \\
Pd$_1$--Pd$_2$&  2.30     &   2.30      & 2.29       &    2.32       \\
Pd$_2$--Pd$_3$&  2.30     &   2.31      & 2.28       &    2.30 \\
Pd$_3$--Pd$_4$&  2.30     &   2.30      & 2.28       &    2.29       \\ 
Pd$_4$--Pd$_5$, Pd$_5$--Pd$_6$ & 2.30 &   2.30      & 2.30       &    2.30     \\ \hline
Adsorption Energy (eV)\hspace{3pt}   &           & 2.841      & 2.4997     &    2.4993   \\ \hline
\end{tabular*}
\end{table}

The adsorption energy per atom is around 2.84~eV for Ag atoms in the first layer,
decreasing to around 2.50~eV for Ag atoms in the second layer and to 2.52~eV in
the third layer. The remarkable fact is that the energy
is virtually the same for the second layer adsorbed
in ``regular'' fcc sites (AB) and in the ``wrong'' hcp sites (AC), i.e. with a stacking fault.
The slight preference of 0.4~meV for the regular structure is in fact smaller than
the expected accuracy of the DFT calculation. This indicates that there is no 
clear preference for either of the two structures on the basis of the energy of the
complete layer alone.

In order to investigate the energetics of the initial nucleation of the second
silver layer, we have calculated configurations in which on top of a 
fcc Pd--Ag structure there is an extra silver atom at a $1/4$ coverage,
either in the fcc or in the hcp threefold hollow site. We
used a $2 \times 2$ supercell along the surface plane, with three sites 
in the last layer empty and one occupied.
We found that the adsorption energy is around 1.90~eV, which
is about 0.6~eV less than the energy per atom in the full silver
monolayer. Due to the large energy difference the silver adatoms 
will tend to aggregate, favouring island formation and layer-by-layer
growth mode. 

The energy of the extra adatom is better by around 0.6~meV if it sits
in the hcp site instead of the fcc site, but this energy difference
is still too small for an unambiguous interpretation.
However, even a slight preference for hcp sites in the initial stages
of the formation of the second Ag layer 
may be sufficient to lock the further growth of the layer 
into the hcp configuration. Once the layer is completed,
the resulting stacking fault (AC order of Ag layers) is frozen since there 
is no simple kinetic path which converts the whole layer into the fcc 
structure (AB). The third Ag layer then starts growing in fcc sites with 
respect to the first and second layer, which results in a ACB layer structure.
This ordering remains, although according to our calculations a three layer
Ag film with ABC structure (which continues the Pd stacking order) 
would have by 7.5~meV per atom better energy. Further silver 
layers grow in the fcc order with respect to the first and second Ag layer,
which results in a twinned crystal structure at the Pd--Ag boundary. 

Analogous formation of a stacking fault between the first and the second 
adlayer has been observed during Ag growth on Pt(111) at room temperature,
but the second layer reverts to the fcc sites after annealing to 
750~K~\cite{Rangelov95}. Further Ag layers continue to grow in a fcc
stacking order with respect to the first two Ag layers, i.e. with or
without a stacking fault depending on whether the second layer was
annealed. The stacking order of thicker Ag films on Pt(111) is not changed by annealing.

We must, however, stress that the energy differences between various configurations 
obtained in our \textit{ab initio} calculations of Ag on Pd(111) are very small, 
and cannot be regarded as a definite explanation of the growth mode.
It has been suggested that the occurrence of a stacking fault between the
first and the second silver layer in the case of Ag growth on Pt(111) is a 
consequence of the kinetics of the adlayer growth rather than of a more 
favourable energy~\cite{Rangelov95}.

When investigating the stability of (111) surface of Au, Pt, and 
Cu in a previous work~\cite{Crljen03}, we have also noted that an
isolated adatom of the same species which adsorbs 
on a completed layer prefers the hcp rather than the regular fcc site. 
However, a full last layer of these three metals in the regular fcc 
position is more stable than a layer in the hcp position by 10, 30, and 
10~meV per atom, respectively. These rather large energy differences
obviously make irrelevant the preference for the hcp site which exists
for a single adatom, and it can be assumed that the fcc sites 
become more stable at an early stage 
of the formation of next layer.
No stacking faults on these surfaces have been observed.

We have also calculated the energies of 1--3 ML Ag films with some other 
stacking orders, but those appear to have less favourable energies.
In particular, the structure where already a monolayer
of Ag is in hcp sites has an energy larger than that of the 
fcc Ag monolayer by around 8~meV, effectively ruling out this mode of growth.

\section{The electronic bands and the QW states}

In order to understand the electronic structure of ultrathin silver films on
Pd(111), we have first made separate calculations of
the electronic bands of the palladium bulk and of unsupported silver films. 

\subsection{The projection of bulk Pd states onto the Pd(111) surface}

\begin{figure}[htb]
\noindent
\begin{center}
\resizebox{0.99\columnwidth}{!}{
\includegraphics[clip=true]{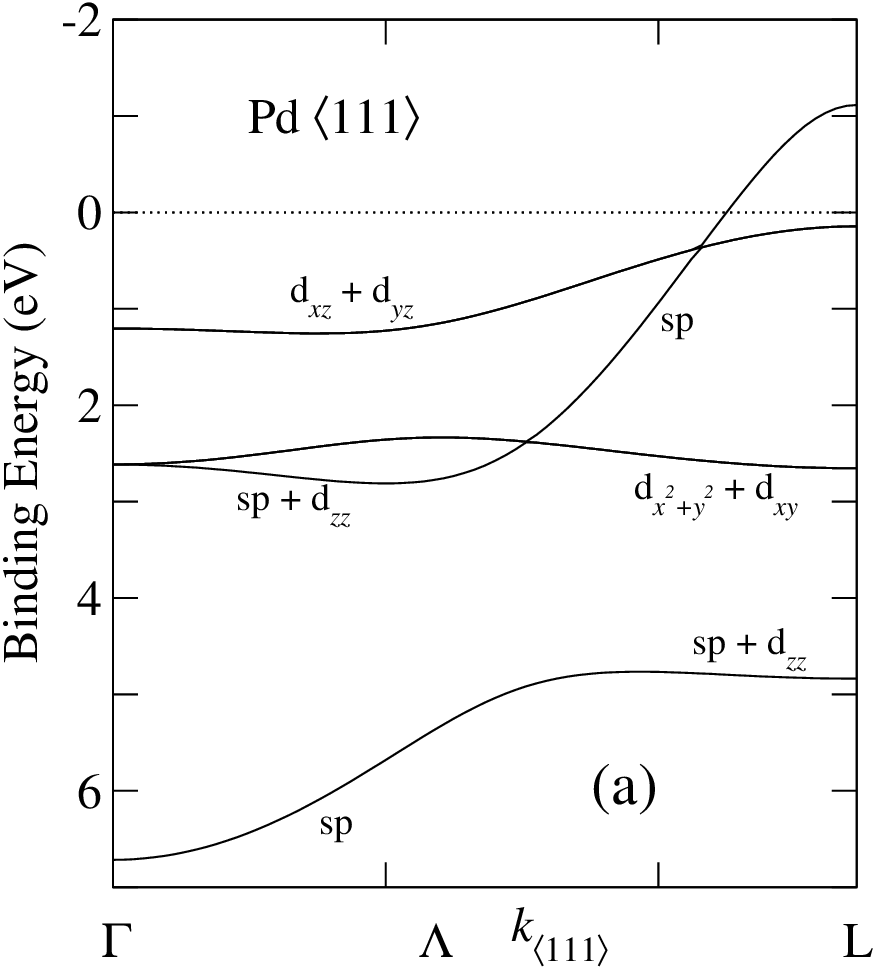}
\hspace{0.5em}
\includegraphics[clip=true]{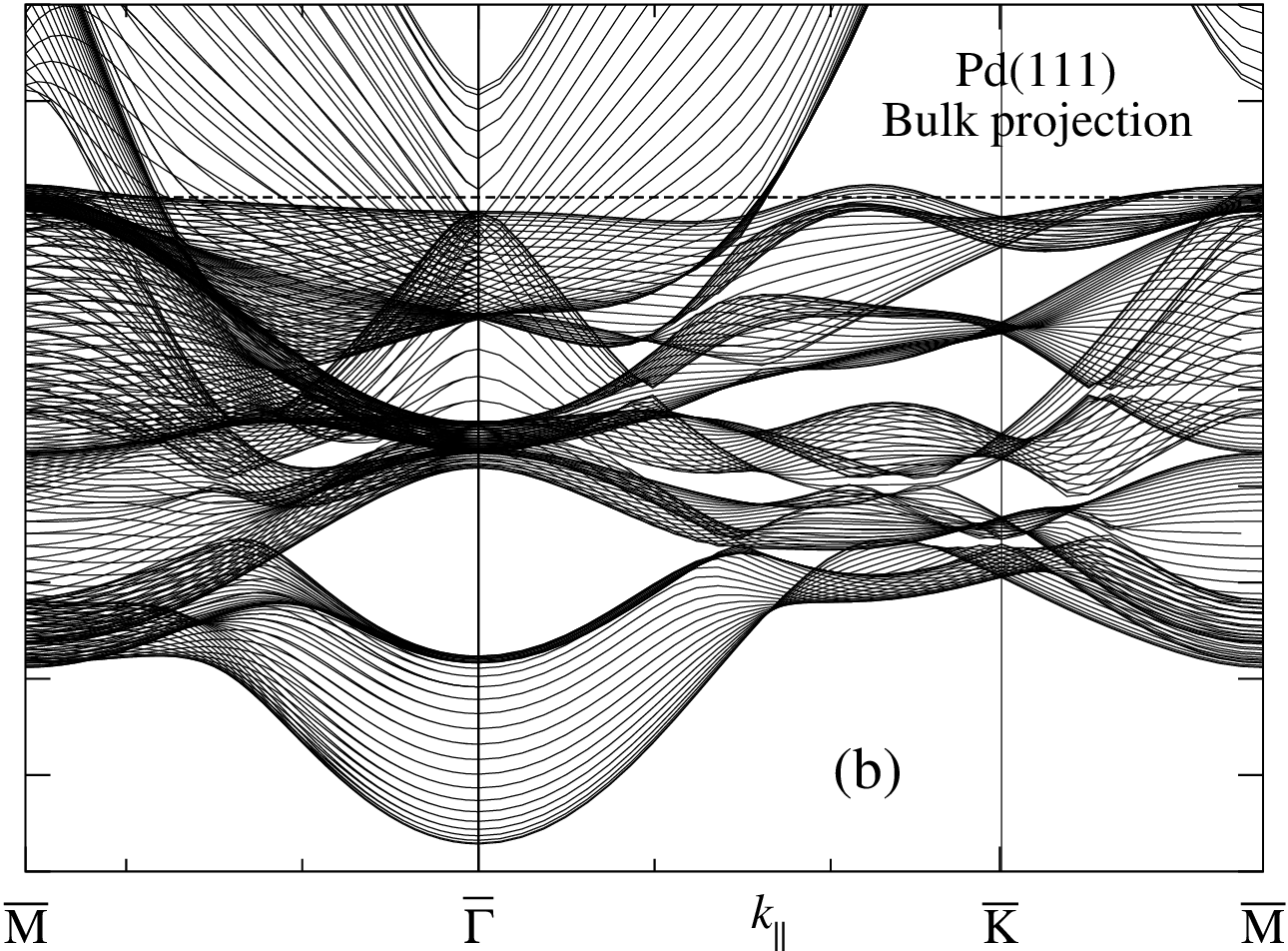}
}
\end{center}
\caption{Calculated electronic structure of bulk palladium. (a) Bulk states of Pd along
$k_z$ direction. (b) Bulk states of Pd projected onto (111) surface.}
\label{bulk}
\end{figure}

In Fig.~\ref{bulk}  we show the calculated electronic structure of bulk palladium 
along the $\left< 111\right> $ direction, 
and  the band structure projected onto the 
surface Brillouin zone (SBZ) for the (111) surface in the 
$\overline{\Gamma}-\overline{\mathrm{M}}$ and $\overline{\Gamma}-\overline{\mathrm{K}}$
directions, for around 30 values of $k_{111}$. At the $\overline{\Gamma}$ point 
of the SBZ the d bands occur from just below the Fermi level to 3 eV, and there is an 
energy gap in the projected states from around 3 to 5 eV binding energy.  

\begin{figure}[htb]
\noindent
\begin{center}
\rotatebox{0}{
\resizebox{0.85\columnwidth}{!}{
\includegraphics[clip=true]{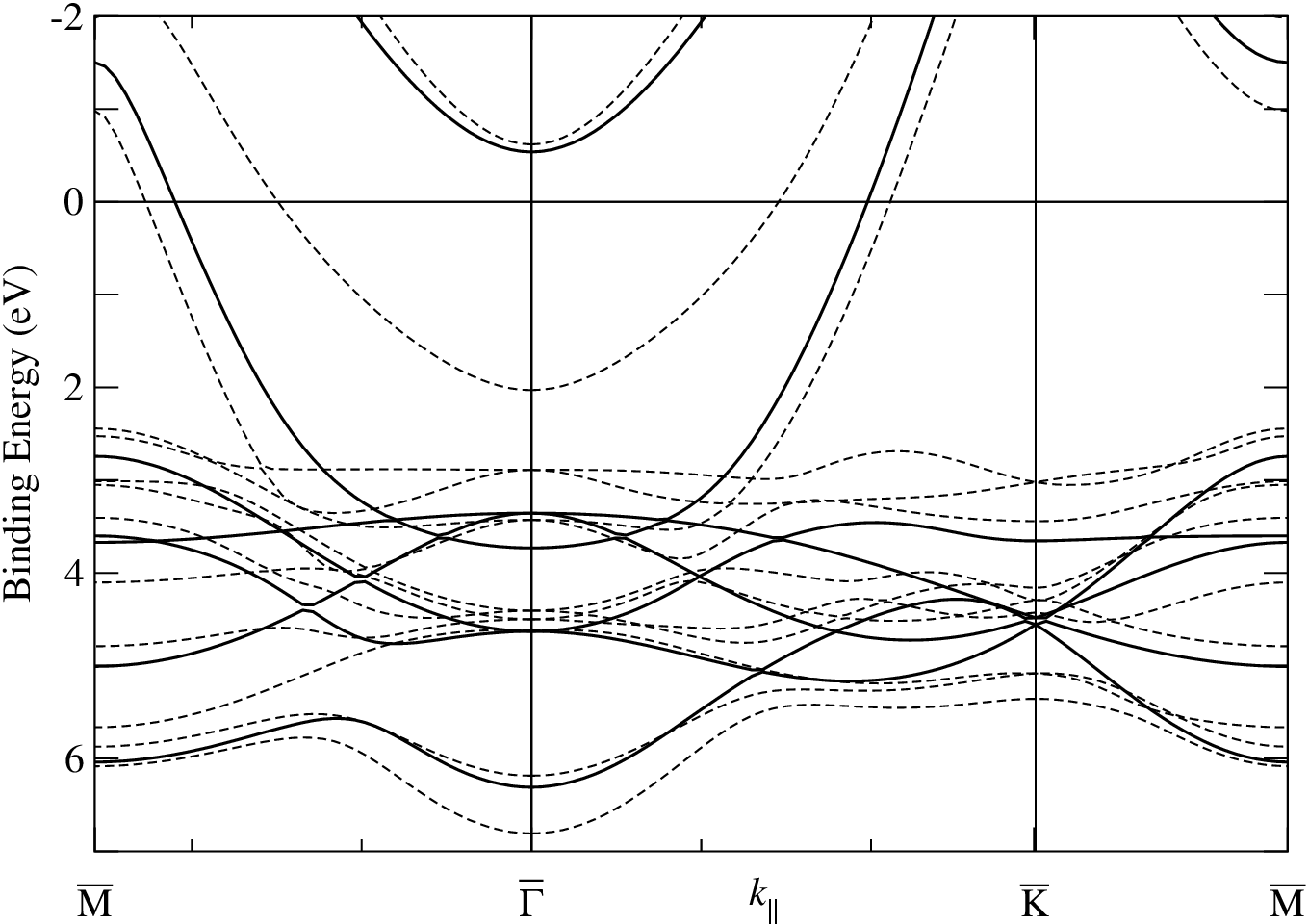}
}}
\caption{Electronic eigenstates of unsupported single (full line)
and double (dashed line) hexagonal layer of Ag atoms, along
the high symmetry directions in the two-dimensional Brillouin zone.}
\label{1-2MLAg}
\end{center}
\end{figure}

\subsection{Unsupported silver films}

Next we take a look at unsupported two-dimensional silver films,
i.e. one or two layer of Ag atoms with vacuum on both sides.
In Fig.~\ref{1-2MLAg} we show the electronic structure of a two-dimensional hexagonal
lattice of Ag atoms with the interatomic distance of 2.3 \AA, corresponding to 
the interatomic separation of Pd(111), along the high symmetry directions of 
the BZ. Narrow d bands are located between 
3.5 and 5.5 eV binding energy, while the broad sp band starts from 6.5 eV 
and crosses the Fermi level near the outer edge of the two-dimensional BZ,
hybridising strongly with the d bands while crossing their energy range.
The electronic structure of a similar two-layer hexagonal structure, with such
interlayer spacing that all distances between the neighbouring silver atoms are
around 2.3 \AA, is shown by dashed lines in Fig.~\ref{1-2MLAg}. The bands look 
similar to the single-layer structure, but their number increases. The 
similarity is particularly evident around the $\overline{\Gamma}$ point, where the
main effect is just the doubling and a small energy offset of the bands.

\begin{figure}[htb]
\noindent
\begin{center}
\rotatebox{0}{
\resizebox{.85\columnwidth}{!}{
\includegraphics[clip=true]{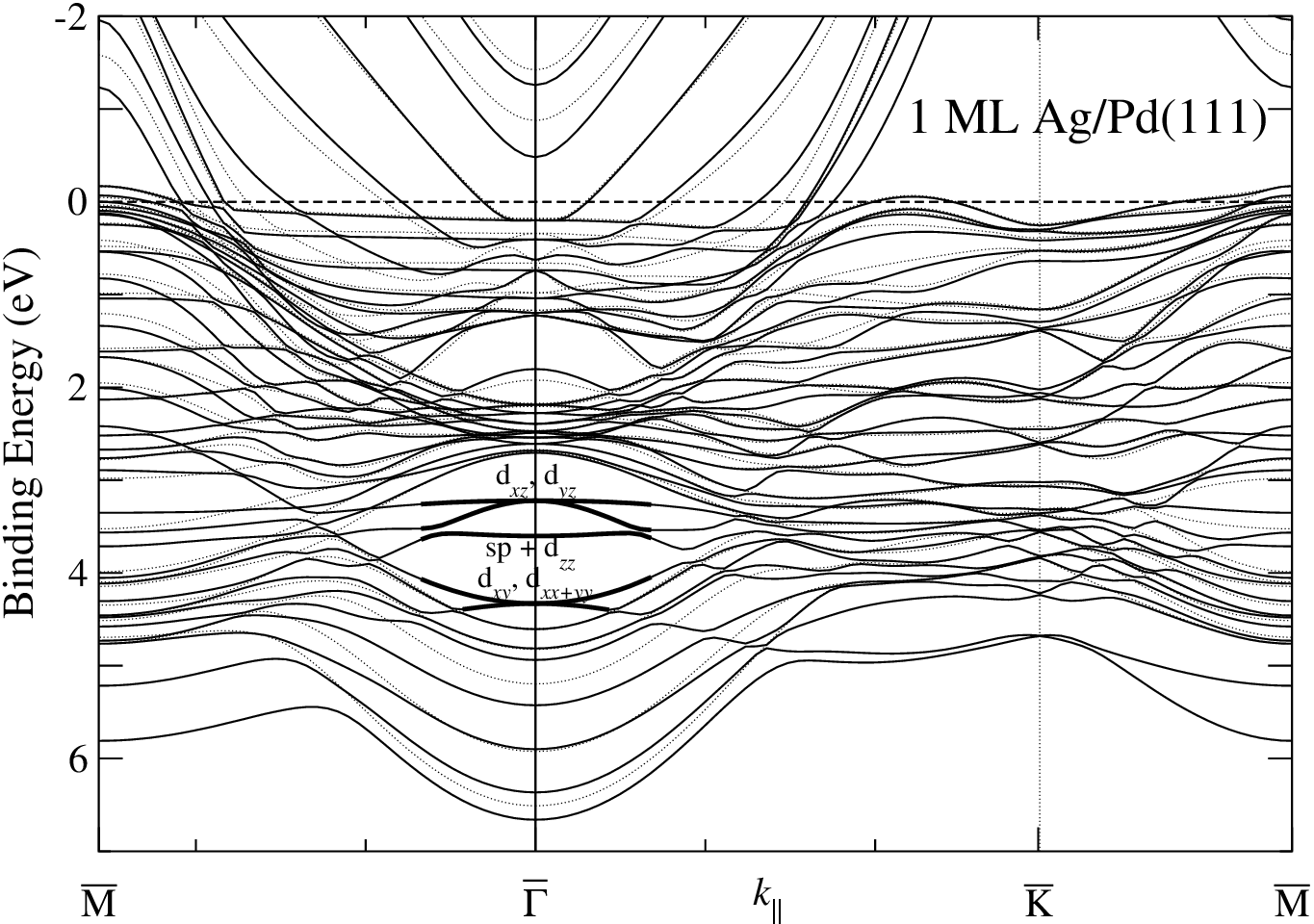}
}}
\caption{Electronic eigenstates of a single layer of Ag atoms
adsorbed on a Pd(111) surface, along
the high symmetry directions in the surface Brillouin zone.
The dotted lines show the states of a clean six-layers Pd substrate.}
\label{1MLAg_Pd}
\end{center}
\end{figure}

\subsection{Ag monolayer on Pd(111)}

In Fig.~\ref{1MLAg_Pd} we show the band structure of a monolayer of 
Ag atoms adsorbed on a Pd(111) surface. The bands of a clean 
six-layer Pd(111) structure are also shown as dotted lines, making it possible to
identify the additional features introduced by the silver layer.
The silver-induced bands which lie in the energy gap of Pd
around the centre of the surface Brillouin zone 
are drawn by thick lines, and their symmetry at $\overline{\Gamma}$
point in terms of atomic orbitals is indicated. 
The absence of propagating (bulk) states in Pd in this $E-k$
region makes the palladium substrate almost as confining 
to the silver electrons
as the vacuum on the other side, so that they
acquire the character of quantum well (QW) states, well
localised within the silver adlayer.
These QW states are the prime candidates to be seen as narrow peaks in 
photoemission experiments in the direction along the normal to the 
surface. 

However,
in calculating the electronic bands of ultrathin silver films on the
V(100) surface, we have found that the silver d states appear about 
1~eV too close to the Fermi level when compared to the peaks 
observed in the photoemission experiments. This deficiency of
the Kohn-Sham eigenstates of d orbitals obtained in density functional
calculations of silver and other noble metals has been known for
some time~\cite{Fuster03}, and we presume that the same will be the case in this
calculation. If we shift the calculated d bands of silver downwards
for around 1~eV, only the bands of $\mathrm{d}_{xz}$,
$\mathrm{d}_{yz}$ symmetry will stay within the energy gap of Pd.
Furthermore, our calculations
does not include the spin-orbit coupling, which leads to a splitting 
of degenerate d bands of silver
clearly observable in photoemission experiments on Ag films on
V(100) surface~\cite{Kralj03b}. 

We can thus expect that the most prominent feature in normal
photoemission from 1~ML Ag/Pd(111) films will come from the QW states
of d character in the region
around 4~eV binding energy. Preliminary experimental
results show that this is indeed the case~\cite{Pervan04}. The normal
PE spectra from silver monolayer on Pd(111), in addition 
to broad maxima centred around 1~eV and 2~eV due to the palladium substrate,
show two narrow peaks just below 4~eV,
interpreted as d QW states separated by a spin-orbit splitting of
around 0.3~eV.

\section{Conclusions}

We have made \textit{ab initio} density functional calculations of the 
structure and electronic properties of ultrathin Ag films 
on Pd(111) surface. We have shown that energetically most favourable 
configuration of the first Ag layer is the one in which the atoms occupy the fcc 
three-fold hollow sites, continuing the substrate structure (with somewhat
different interlayer spacing, of course). The second Ag layer, however,
has almost the same energy whether the atoms sit in fcc or in hcp three-fold
hollow sites. We have found that the first extra atoms to adsorb on a completed 
first layer in fact prefer the hcp site, which makes it likely that this
mode of growth continues up to the completion of the second layer. This is a 
possible explanation of the experimentally observed stacking fault between the 
first and the second Ag layer. We have also calculated the electronic band
structure of Ag/Pd(111) systems. We have found that around the centre of
the two-dimensional surface Brillouin zone of Pd(111) there is an energy
gap in the density of states in the 3--5~eV energy range. Upon adsorption
of a monolayer of Ag, quantum well states of predominantly d character 
form at binding energies around 4~eV.

\section*{Acknowledgements}

This work was supported by the Ministry of Science, Education and Sports 
of the Republic of Croatia under contract No. 0098001.
We are grateful to P.~Pervan and M.~Milun for the communication
of their unpublished results and for useful discussions.

\end{document}